\documentclass[%
 aip,
 pof,%
 reprint,%
]{revtex4-1}
\usepackage{float}
\usepackage{amsmath}
\usepackage{amssymb}
\usepackage{hyperref}
\usepackage[none]{hyphenat}
\usepackage{graphicx}
\usepackage{fixltx2e}
\usepackage{epstopdf} 

\begin{document}

\title{Field-induced shaping of sessile paramagnetic drops}
\author{Jennifer Dodoo}
\affiliation{School of Engineering, Institute for Integrated Micro and Nano Systems, The University of Edinburgh, Edinburgh, EH9 3LJ, United Kingdom}
\author{Adam A. Stokes}
\email[corresponding author: ]{adam.stokes@ed.ac.uk} 
\affiliation{School of Engineering, Institute for Integrated Micro and Nano Systems, The University of Edinburgh, Edinburgh, EH9 3LJ, United Kingdom}

\date{\today}

\begin{abstract}
We use the electromagnetic stress tensor to describe the elongation of paramagnetic drops in uniform magnetic fields. This approach implies a linear relationship between the shape of the drops and the square of the applied field which we confirm experimentally. We show that this effect scales with the volume and susceptibility of the drops. By using this unified electromagnetic approach, we highlight the potential applications of combining electric and magnetic techniques for controlled shaping of drops in liquid displays, liquid lenses, and chemical mixing of drops in microfluidics.
\end{abstract}
\maketitle
\begin{figure}[b]
\includegraphics[width = 1\linewidth]{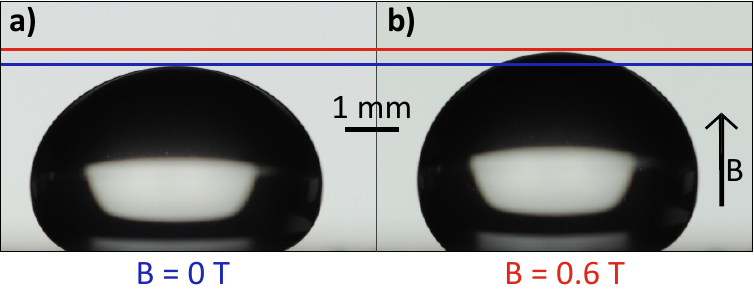}
\caption{\label{fig1} Images of a 60 $\mathrm{\mu}$l drop of an aqueous solution with 51.4\% ppw $\mathrm{GdCl_3\cdot6H_2O}$ \textbf{a)} in the absence and \textbf{b)} in the presence of a magnetic field. The horizontal lines highlight the height of the drop in the absence (blue) and presence (red) of the applied field.}
\end{figure}
\begin{figure}[t]
\includegraphics[width = 1\linewidth]{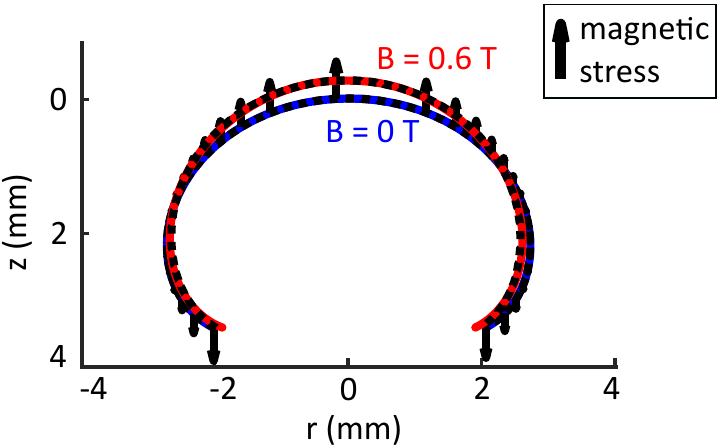}
\caption{\label{fig2} The outlines (black dashed) of the drop in Fig. \ref{fig1}, and the corresponding numerical solutions of Eq. (\ref{eqn:YL_em}). The black arrows show the variation of the relative magnitude of the magnetic stress along the outline of the drop.}
\end{figure}

Controlled shaping of small volumes of fluids (or drops) is a key ingredient for liquid lenses to set optical properties \cite{Mishra2016}, and for digital microfluidics (DMF) \cite{Seemann2011,Samiei2016,Freire2016a,Ding2019}, where drops are, for example, manipulated for chemical mixing \cite{Chen2019}. Drops can be shaped through the application of electromagnetic fields which exert a force on ions, and electric and magnetic dipoles in the drop.\\
Electric actuation techniques for DMF include electrowetting \cite{Mugele2005a,Teng2020} - a technique where surface energies of the substrate are electrostatically modified, which requires contact between the electrodes and the drop - and liquid dielectrophoresis \cite{McHale2011} - a contact-free bulk effect, where a non-uniform electric field is applied to electric dipoles. Electrowetting is widely implemented in DMF devices \cite{Seemann2011,Samiei2016,Ding2019,Teng2020} and liquid lenses \cite{Mishra2016}. Dielectrowetting has been explored for liquid optics \cite{Brown2009} and has been successfully implemented for DMF manipulations \cite{Geng2017}. More recently, electronic dewetting using ionic-surfactants has been explored \cite{Li2019}.\\
In contrast to electric actuation techniques stand magnetic actuation techniques, which include: liquid marbles where magnetic particles are used as a coating \cite{Zhao2012} or are inserted into the marble \cite{Bormashenko2008,Khaw2016}; the actuation of fluids through the deformation of magnetic substrates \cite{Chen2019}; or the actuation of magnetic particle suspensions, such as ferrofluids \cite{Pamme2006,Zhang2017}. Ferrofluidic drops in uniform magnetic fields have been studied in suspension \cite{Afkhami2010} and on superhydrophobic surfaces \cite{Zhu2011}. Ferrofluidic drops are commonly used in DMF, because they can be actuated at low field strength due to their high magnetic susceptibility values ($\chi>10,000$) \cite{Stierstadt2015, Latikka2018}. Drops of ferrofluids in magnetic fields form cones and spikes once a threshold of applied field strength is reached \cite{Latikka2018}.\\
An alternative to ferrofluids are paramagnetic salt solutions, which contain uniformly distributed, randomly oriented, weak magnetic dipoles, and have therefore a much smaller $\chi$ ($\mathrm{<<1}$) than ferrofluids. The potential application of this alternative 'particle-free' actuation method to DMF has been demonstrated by actuating drops containing various paramagnetic salt solutions on superhydrophobic surfaces where a strong correlation between $\chi$ and ease of actuation has been shown \cite{Egatz-Gomez2006, Mats2016} and implemented for electrochemical detection \cite{Lindsay2007} and fluorescence measurements \cite{Mats2016}.\\
A common approach to describe the shape of drops is by balancing the differences in stresses, such as stresses due to gravity and surface tension, across the liquid-vapour interface (Young-Laplace equation). The Young-Laplace equation can be adapted for electromagnetic fields by including in the stress balance the difference in electromagnetic stress - usually evaluated using the Maxwell stress tensor (MST). The Young-Laplace equation has been modified accordingly and applied to the deformation of drops in electromagnetic fields, including the break-up of drops in electric (magnetic) fields \cite{Sherwood1988}; the strong deformation and formation of conical ends of dielectric drops in electric fields \cite{Basaran1992,Stone1999}; the deformation of conducting drops in uniform electric fields \cite{Corson2014}; the instabilities of ferrofluidic drops in magnetic fields \cite{Bacri1982,Stone1999}; and the dynamics of the deformation of suspended ferrofluidic drops\cite{Afkhami2010,Rowghanian2016}.
The MST is however valid in vacuum, as it is a reduced form of the electromagnetic stress tensor (EMST), which is universally valid for quasi-static non-dissipative processes \cite{Stierstadt2015}. The EMST has recently been successfully applied to suspended ferrofluid drops \cite{Rowghanian2016}. The EMST generally depends on: the thermodynamic potential of the system; the electric permittivity; the magnetic susceptibility; and the electromagnetic field applied.\\
We have recently validated a modified Young-Laplace equation experimentally for the field-induced change in shape of diamagnetic sessile drops \cite{Dodoo2019a}. Here, we generalize this approach further, by explicitly deriving the modified Young-Laplace equation and numerically fitting it to the outline of sessile paramagnetic drops in uniform magnetic fields. Our results show that (i) the field-induced change in shape of paramagnetic drops is due to the magnetic stress difference across the liquid-vapour interface; (ii) the deformation increases proportionally with the square of the applied field strength; (iii) drops with a large volume and magnetic susceptibility elongate more than drops with a small volume and magnetic susceptibility.\\
The equilibrium shape of a sessile drop is determined by the stresses acting on it; these may include but are not limited to interfacial, gravitational and electromagnetic stress. To derive an expression for the shape of a sessile drop, we follow an analogous method to the one presented by Stierstadt and Liu \cite{Stierstadt2015} and use their definition of the full electromagnetic stress tensor which is universally valid for time-independent (quasi-static) non-dissipative processes:
\begin{equation}
\footnotesize{\sigma_{ik} = U - TS - \xi_{\alpha} \rho_{\alpha} - \mathbf{E}\cdot\mathbf{D}-\mathbf{H} \cdot \mathbf{B})\delta_{ik}+E_iD_k+H_iB_k}
\label{eqn:EMST}	
\end{equation}
where \textit{i} is the direction of force and \textit{k} is the direction normal to the surface to which the force is applied, $U$ is the total energy density of matter and field (J m$^{-3}$), $T$ is temperature (K), $S$ is the entropy of matter and field (J m$^{-3}$ K$^{-1}$), $\xi_{\alpha}$ is the mass density of the chemical potential (J kg$^{-1}$) of the material component $\alpha$, $\rho_{\alpha}$ is the partial density of the material component $\alpha$ (the total density is $\rho^{tot}=\Sigma_{\alpha}\rho_{\alpha}$), $\mathbf{E}$ is the electric field strength, $\mathbf{D}$ is the electrical displacement, $\mathbf{H}$ is the auxiliary field, $\mathbf{B}$ is the magnetic flux density, and $\delta_{ik}$ is the Kronecker-Delta function.\\
In this work, we assume that there are no electric fields acting on the drop ($\mathbf{D}$=$\mathbf{E}$=0) and that the magnetic susceptibility is independent of the applied magnetic field strength ($\mathbf{B} = \mu \mathbf{H} = \mu_0 (1+\chi)\mathbf{H}$ and $\mathbf{M} = \chi\mathbf{H}$).  We limit this analysis to a closed thermodynamic system at constant temperature and volume, which is suitably described by the Helmholtz potential ($U = a^{t}+TS$). The thermodynamic potential can be separated into field-independent ($a_0$) and field-dependent ($a_{em}$) terms: $a^{t}=a_0+a_{em}$, with $a_{em}=\int \mathbf{H}\cdot \mathbf{dB}$.\\
The definition of the mass density of the chemical potential is: $\xi_{\alpha}=\delta a^t/\delta \rho_{\alpha}$ \cite{Stierstadt2015}. In equilibrium, the stresses on the boundary of the two substances from inside and outside must be equal. The vapour phase is air, which we approximate as vacuum in its magnetic properties ($\mu=\mu_0$) as well as in its chemical potential ($\xi_{\alpha}\rho_{\alpha}=0$). We impose the standard boundary conditions for Maxwell's equations as formulated by Stierstadt and Liu \cite{Stierstadt2015}: (1) the difference in the magnetic field component tangential to the surface must vanish ($\Delta B_t=\Delta H_t=0$); (2) the stress components normal to the interface, $\sigma_{nn}$, must be continuous, while tangential components cancel. The electromagnetic stress difference across the air-liquid boundary then becomes:
\begin{equation}
\footnotesize{\Delta\sigma_{nn}^{EM}=a_0^l-a_0^v-\xi_0\rho_{\alpha}-\frac{\mu_0}{2} H^2 \bigg(\chi+\rho_{\alpha}\frac{\delta \chi}{\delta\rho_{\alpha}}\bigg)+\mu_0\chi H_n^2}
\label{eq:em_general}
\end{equation}
The magnetic flux density is the vector sum of its normal and tangential components with respect to the surface over which $\Delta\sigma_{nn}^{EM}$ is resolved, $H^2=H_n^2+H_t^2$. 
We calculate the chemical potential using the Clausius-Mossotti approach: $\rho (\delta\chi/\delta\rho) = \chi(1+\chi/3)\approx\chi$ for one-component fluids with $\chi<<1$. 
\begin{equation}
\Delta\sigma_{nn}^{EM}=a_0^l-a_0^v-\xi_0\rho+\frac{\chi}{\mu_0} \bigg(B_n^2-B^2\bigg)
\label{eq:em-stress}
\end{equation}
where B is the magnetic flux density in air (T). In addition to the electromagnetic stress difference, the shape of the drop is also determined by the stress differences due to gravity and surface tension\cite{Stierstadt2015,2008Miller}:
\begin{align}
\Delta\sigma_{nn}^{surf} &= \gamma (R_1^{-1}+R_2^{-1})\\
\Delta\sigma_{nn}^{grav} &= g \Delta\rho z
\label{eqn:sigma_grav}
\end{align}
where $\gamma$ is the surface tension (N m$^{-1}$), $R_1$ and $R_2$ are the principle radii of curvature of the drop outline (m), $g$ is the gravitational acceleration (m s$^{-2}$), $\rho$ is the mass density (kg m$^{-3}$), and $z$ is the vertical distance to the apex point (m), as indicated by Fig. \ref{fig2}. The principle radii of curvature of an axisymmetric drop are commonly expressed as  \cite{Sherwood1988,Adamson1990,2008Miller,Lubarda2011}:
\begin{equation}
\frac{1}{R_1}+\frac{1}{R_2} = \frac{r''}{(1+r'^2)^{3/2}}-\frac{1}{r(1+r'^2)^{1/2}}
\end{equation}
where r(z) is the drop outline, originating from the apex point, $r'$ and $r''$ are the first and second derivatives of r with respect to $z$. 
In equilibrium, the stresses on the drop must sum up to zero. 
\begin{equation}
0 = \Delta\sigma_{nn}^{surf} + \Delta\sigma_{nn}^{grav} + \Delta\sigma_{nn}^{EM} \label{eqn:YL_em}
\end{equation}
Eq. (\ref{eqn:YL_em}) is an augmented Young-Laplace equation, which describes the hydrostatic and magnetic stresses on a sessile magnetic drop in air due to a magnetic field. Note that the definition of the magnetic stress in Eq. (\ref{eq:em_general}) holds for ferrofluids and para- and diamagnetic salt solutions and is valid in any time-independent (static) magnetic field. Rowghanian \textit{et al.} have derived this modified Young-Laplace equation from the EMST using dimensionless parameters \cite{Rowghanian2016}. 
For axisymmetric drops, the radius of curvature at the apex point, b, depends on the difference of the energy densities of the Helmholtz potential of the liquid and vapour phase\cite{Stierstadt2015} $2\gamma b^{-1}=a_0^l-a_0^v$. In the absence of magnetic fields, $\mathbf{B}=\mathbf{H}=0$, the stress balance of the drop simplifies to $0=(R_1^{-1}+R_2^{-1})+2 b^{-1}+(g\Delta\rho\gamma^{-1})z$, which is the well-known Young-Laplace equation \cite{Adamson1990,2008Miller,Lubarda2011}.\\
To test the validity of Eq. (\ref{eqn:YL_em}), we experimentally investigate the shape of paramagnetic drops in a homogeneous magnetic field directed along the symmetry axis of the drop (Fig. \ref{fig1}). We use a C-frame adjustable electromagnet containing iron cores with tips of 8 mm diameter. The substrates are 1 mm thick microscope glass slides coated with superhydrophobic composite films from colloidal graphite \cite{Bayer2012}. To measure over a wide range of total magnetic moments ($J=\chi V B \mu_0^{-1}$), we apply fields of 0 to 0.6 T; use drop volumes of 40-100 $\mu$l; and use three different aqueous paramagnetic salt solutions: (1) 17.8\% ppw ($\chi=1.56\times10^{-4}$) and (2) 35.6\% ppw ($\chi=3.21\times10^{-4}$) manganese chloride tetrahydrate ($\mathrm{MnCl_2\cdot 4H_2O}$) and (3) 51.4\% ppw ($\chi=5.24\times10^{-4}$) gadolinium chloride hexahydrate ($\mathrm{GdCl_3\cdot 6H_2O}$) \footnote{
We calculate $\chi$ for the salt solution using: $\chi = (\chi_s^m C_s M_s^{-1}+\chi_w^m (1-C_s) M_w^{-1})\rho$, where $M_s=197.9$ g mol$^{-1}$ for $\mathrm{MnCl_2\cdot 4H_2O}$, and $M_s=371.7$ g mol$^{-1}$ for $\mathrm{GdCl_3\cdot 6H_2O}$, and $M_w=18.02$ g mol$^{-1}$ are the molecular masses, and $\chi_s^m=14350\times10^{-6}$ cm$^3$ mol$^{-1}$ for $\mathrm{MnCl_2\cdot 4H_2O}$, and $\chi_s^m=27930\times10^{-6}$ cm$^3$ mol$^{-1}$ for $\mathrm{GdCl_3\cdot 6H_2O}$, and $\chi_w^m=-12.63\times10^{-6}$ cm$^3$ mol$^{-1}$ are the literature values for the molar magnetic susceptibilities of the paramagnetic salts and water respectively \cite{CRCinorganic}. $C_{s}$ and $\rho$ are the weight concentration of the salt in the solution and the density of the solution respectively.}.\\
To analyse the shape of the drop, we developed an algorithm similar to the standard Axisymmetric Drop Shape Analysis in electric fields \cite{Bateni2004,Saad2016}: by imaging the side-profile of the drop using a digital DSLR-camera and using computational image analysis, we determine the drop outline r(z); we solve Eq. (\ref{eqn:YL_em}) numerically for r(z), using the Runge-Kutta method; and iteratively fit the solution of Eq. (\ref{eqn:YL_em}) using a standard least-square method (Levenberg-Marquardt) to the left and right-side of the drop outline independently.\\
To obtain good fits of Eq. (\ref{eqn:YL_em}) to r(z) of (1) the drop in the absence of magnetic fields: we estimate the density of the solutions $\rho$ to be equal to the density of water ($\rho_w=997$ kg m$^{-3}$) and 1.1 times the density of water for the $\mathrm{MnCl_2\cdot 4H_2O}$ and $\mathrm{GdCl_3\cdot 6H_2O}$ solutions respectively and allow $\gamma$ to freely change; and (2) the drop in the presence of magnetic fields: we allow $\xi_0\rho_\alpha$ to freely change. This numerical optimization of physical values allows us to account for errors in our estimates of (1) the surface tension and density of the drop; and (2) the value of the field-independent chemical potential and the value of the difference of the field-independent thermodynamic potentials of liquid and vapour phase ($a_0^l-a_0^v$). The uncertainty on the $a_0^l-a_0^v$ value is caused by a non-axisymmetric deformation of the drop, due to systematic errors such as inhomogeneities in the applied field and in the roughness of the substrate, and the coarseness of the manual levelling of the substrate and magnet. (Without a numerical optimization of $\xi_0\rho_\alpha$ (set to zero) we achieve less consistently good fits.)
To measure the radius of curvature at the apex point, we fit a parabolic function to r(z) in the range where Eq. (\ref{eqn:YL_em}) vanishes.\\
An example result of this methodology is presented by Fig. \ref{fig1}. The side-profile photographs of a 60 $\mathrm{\mu}$l drop of the aqueous 51.4\% ppw $\mathrm{GdCl_3\cdot6H_2O}$ solution presented by Fig. \ref{fig1} show a visible elongation of the drop along the field lines. As the drop elongates along the field lines, the width decreases from 5.4 to 5.2 mm and the height increases from 3.4 to 3.7 mm.\\
The numerical solutions of Eq. (\ref{eqn:YL_em}) run smoothly along the outline as shown by Fig. \ref{fig2}, with the optimized numerical value $\gamma=$66.7 mN m$^{-1}$. The optimized surface tension is $\approx8\%$ smaller than that of water (72.8 mN m$^{-1}$), accounting errors in the initial guesses of the numerical values of density and surface tension. The diameter of the triple contact line decreases by 0.05 mm. The fit routine optimizes the change in diameter of the triple contact line to 0.3 mm. The motion of the triple contact line is inhibited by the friction of the surface. This factor has been omitted in the fit routine, leading to the mismatch of real and optimized value. Over the range of investigated salt concentrations, the optimized surface tension values are (72.3$\pm$0.1) mN m$^{-1}$, (71.3$\pm$0.3) mN m$^{-1}$, and (69.8$\pm$0.5) mN m$^{-1}$ for the 17.8\% ppw $\mathrm{MnCl_2\cdot 4H_2O}$, 35.6\% ppw $\mathrm{MnCl_2\cdot 4H_2O}$, and 51.4\% ppw $\mathrm{GdCl_3\cdot 6H_2O}$ solutions respectively and the field-independent chemical potential varies, independently of salt concentration, between -(1.6$\pm0.2$)$\times10^{-3}$ J kg$^{-1}$.
Noise on the measurements originates from physical sources (vibrations induced in the drop from the laboratory environment), and from the grayscale to binary image conversion which introduces a random error caused by the background light and the pixel resolution of the camera. The deformation is fully reversible as long as the volume of the drop remains constant for the duration of the measurement.\\ 
Also presented by Fig. \ref{fig2} is the magnetic stress acting on the outline of the drop. We observe that the stress is directed outwards from the drop along the magnetic field lines and its magnitude is proportional to the normal component of the magnetic flux density $B_n^2$. The magnetic stress is largest at the apex point, where the surface normal of the drop is parallel to the magnetic field lines, and diminishes at the outermost sides of the drop, before increasing again. At the solid-liquid interface, the surface vector of the drop is also parallel to the magnetic field lines, resulting in a magnetic stress and subsequent elongation of the drop towards the solid substrate which has not been studied here. This effect can be observed when suspending the droplet in a non-magnetizable medium \cite{Afkhami2010,Rowghanian2016}. Our investigation demonstrates the axisymmetric deformation of a paramagnetic drop through the application of a magnetic field that is axisymmetric with respect to the drop. A field that is not axisymmetric with respect to the drop, would cause a deformation towards the region of higher magnetic flux density.\\
\begin{figure}[t]
\includegraphics[]{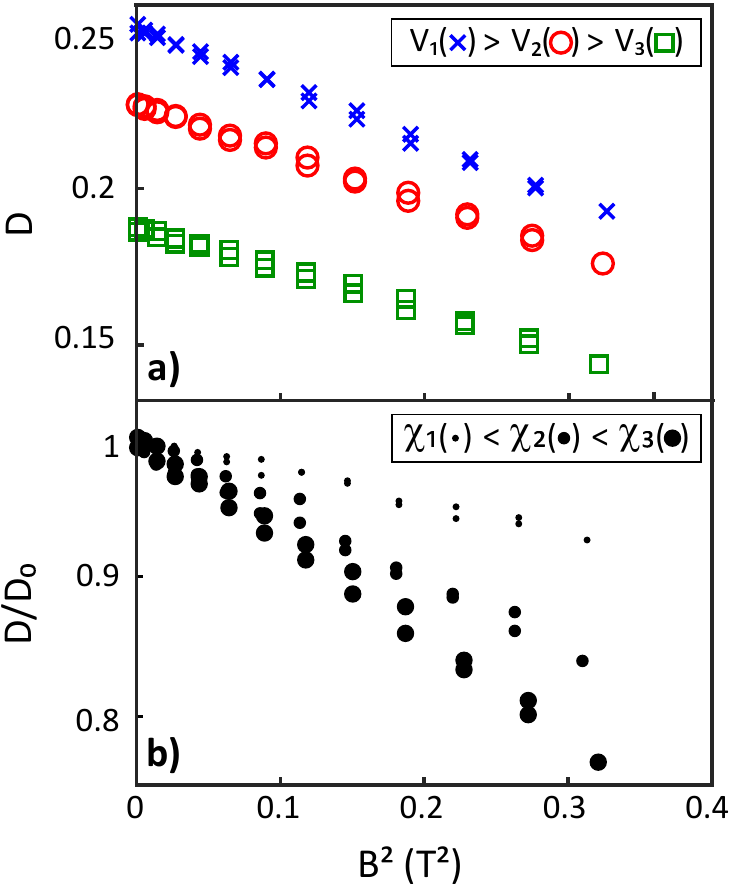}
\caption{\label{fig3}Dimensionless shape parameter D (see text) of drops with different volumes and magnetic susceptibilities shown as a function of applied field squared. \textbf{a)} D of 51.4\% ppw $\mathrm{GdCl_3\cdot 6H_2O}$ drops with volumes $V_{1,2,3}=(80,60,40)$ $\mathrm{\mu}$l. \textbf{b)} D normalized with respect to the drop shape in absence of a magnetic field and shown for $40$ $\mathrm{\mu}$l drops with $\chi_{1,2,3}=(1.56,3.21,5.24)\times10^{-4}$.}
\end{figure}
\begin{figure}[t]
\includegraphics[]{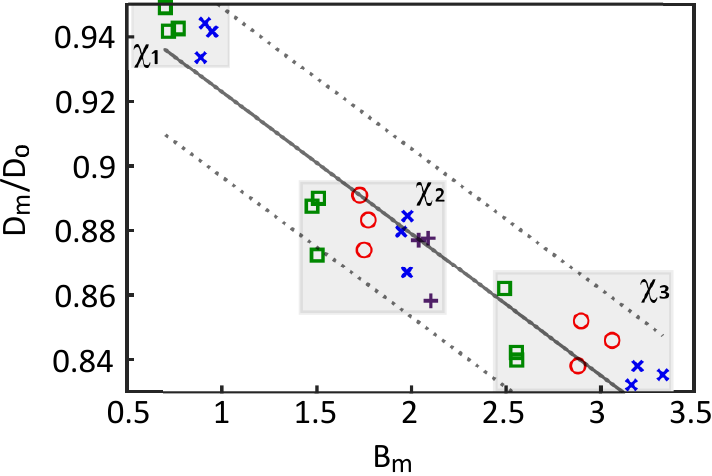}
\caption{\label{fig4}Normalized dimensionless shape parameter in an applied field of $B=0.5$ T ($D_m$) normalized with respect to the drop shape in the absence of a magnetic field ($D_0$) against magnetic bond number $B_m$ (see text). The drop volumes are indicated by the symbols as in Fig. \ref{fig3}a), with the addition of the red cross symbol for $V=100$ $\mathrm{\mu}$l. The $\chi$ values are the same as in Fig. \ref{fig3}b). The line of best fit (solid) and the 95\% prediction interval (dotted) are also shown.}
\end{figure}
Fig. \ref{fig3} shows the change in the dimensionless shape parameter $D=(w-h)/(w+h)$, where $w$ is the width and $h$ is the height of the drop. Fig. \ref{fig3}a) shows $D$ of drops of 51.4\% ppw $\mathrm{GdCl_3\cdot 6H_2O}$ with different volumes against applied magnetic field squared. As the drops elongate along the field lines, $D$ decreases proportionally to $B^2$. The total change in $D$ increases with the volume of the drop.
Fig. \ref{fig3}b) presents $D$ normalized with respect to its value in the absence of a magnetic field ($D/D_0$). The linear decrease of $D/D_0$ in a magnetic field is shown for $40\mu l$ large drops with different magnetic susceptibilities. The total change in $D/D_0$ increases with the magnetic susceptibility.\\
Fig. \ref{fig4} shows the normalized dimensionless shape parameter in the presence of 0.5 T strong magnetic field ($D_m/D_0$) against the magnetic Bond number ($B_m$). The magnetic Bond number is the ratio of the magnetic to surface energy\cite{Nguyen2013}: $B_m=\chi V^{1/3} B^2 (2\gamma\mu_0)^{-1}$. For the calculation of $B_m$ we use the optimized value of $\gamma$. $D_m/D_0$ decreases linearly with $B_m$, where the fitted relationship is: $D_m/D_0=-0.44 B_m +0.97$. The mean error on the fit is $\Delta=0.01$ and the 95\% prediction interval includes $2\Delta$. Drops with a large volume and value of magnetic susceptibility elongate more strongly than small drops that are less susceptible to magnetic fields.\\
In conclusion, paramagnetic drops elongate in uniform magnetic fields due to a mismatch in magnetic stress across the liquid-vapour interface. We investigated this effect (i) analytically using the electromagnetic stress tensor; and (ii) experimentally by analysing the shapes of paramagnetic drops in uniform magnetic fields. We found a linear relation between the shape of the drop and the square of the applied field, which scales with the volume and susceptibility of the drop. Though independent treatments of electric and magnetic phenomena are widely used to describe drop actuation in DMF, such as the Lippmann-Young equation for electrowetting \cite{Mugele2005a}, a unifying electromagnetic treatment highlights the symmetries between electric and magnetic actuation techniques with the potential to increase the versatility of their implementation.\\

The data that support the findings of this study are available from the corresponding author upon reasonable request.

\section*{Acknowledgements}
We thank Glen McHale and Martin Brinkmann for useful discussions. The work by JD was funded by the EPSRC Centre for Doctoral Training in Integrative Sensing and Measurement: EP/L016753/1.
\section*{References}
\bibliography{library}

\end{document}